\documentclass[
 reprint,
showpacs,
 amsmath,amssymb,
 aps,
prb,
]{revtex4-1}

\usepackage{graphicx}
\usepackage{dcolumn}
\usepackage{bm}
\usepackage{hyperref}

\def\v#1{{\bf#1}}
\def\be{\begin{equation}}
\def\ee{\end{equation}}
\def\bea{\begin{eqnarray}}
\def\eea{\end{eqnarray}}
\def\ahalf{{\textstyle{1\over2}}}

\newcommand{\bfalpha}{\mbox{\boldmath$\alpha$\unboldmath}}

\newcommand{\bfsigma}{\mbox{\boldmath$\sigma$\unboldmath}}

\newcommand{\bfgamma}{\mbox{\boldmath$\gamma$\unboldmath}}

\newcommand{\bfkappa}{\mbox{\boldmath$\kappa$\unboldmath}}

\def\hcal{\mbox{$\cal H\,$}}
\def\<{\langle}
\def\>{\rangle}

\begin{document}


\title{Discrete symmetry in graphene: the Dirac equation and beyond}

\author{E. Sadurn\'i}
 \email{sadurni@ifuap.buap.mx}
\author{E. Rivera-Moci\~nos}
\email{erivera@ifuap.buap.mx}
\author{A. Rosado}
\email{rosado@ifuap.buap.mx}
\affiliation{Instituto de F\'isica, Benem\'erita Universidad Aut\'onoma de Puebla,
Apartado Postal J-48, 72570 Puebla, M\'exico}

\date{\today}

\begin{abstract}
In this pedagogical paper we review the discrete symmetries of the Dirac equation using elementary tools, but in a comparative order: the usual $3+1$ dimensional case and the $2+1$ dimensional case. Motivated by new applications of the 2d Dirac equation in condensed matter, we further analyze the discrete symmetries of a full tight-binding model in hexagonal lattices without conical approximations. We touch upon an effective CPT symmetry breaking that occurs when deformations and second-neighbor corrections are considered.
\end{abstract}



\maketitle


\section{Introduction}

The rise of two-dimensional materials and a subsequent avalanche of studies \cite{goldberg2010, katsnelson2007, geim2007} have led to significant advances in theory and experiments. With this, the Dirac equation has found happy applications in electronic transport \cite{dassarma2011}, photonic structures \cite{haldane2008, raghu2008} and recently, ultracold matter in optical lattices \cite{uehlinger2013}.

The crossover between crystalline structures and relativistic quantum mechanics compells us to analyze these systems from different angles. In this paper we are interested in discrete symmetries, whose implications in elementary particle physics have been clearly established and -- in the frontiers of our knowledge -- occasionally tested \cite{ellis1994, maiani1995, abouzaid2011, beringer2012}. 

Our tasks imply a revision of dimensionality and its consequences. The $2+1$ dimensional Dirac equation shares many features with the usual $3+1$ dimensional case, but there are also differences that manifest themselves in discrete transformations and the nature of chiral symmetries. In a more general framework, we should point out that nearest-neighbor tight-binding models allow exact solutions, and that their formulation goes beyond the Dirac approximation. Therefore, this is an excellent opportunity to discuss discrete symmetries in a more general setting. As a bonus, we shall see that a symmetry breaking analogous to CPT violation may occur beyond effective Dirac theories.

We present our discussion in the following order: In section II we provide the concepts that explain the appearence of discrete transformations as members of the Lorentz group. We also review the origin of the Dirac equation and show how its spinorial dimensionality is related to space-time dimensionality. In section III we focus on parity, analyzing both $3+1$ and $2+1$ dimensional cases. Section IV is devoted to effective Dirac theories; in this section we study the effects of parity on hexagonal lattices and suggest a symmetry breaking of full tight-binding models. We conclude in section V. 

\section{Preliminary concepts}

\subsection{The sheets of the Lorentz group}

\begin{figure}[t]
\begin{tabular}{c}  \includegraphics[width=4cm]{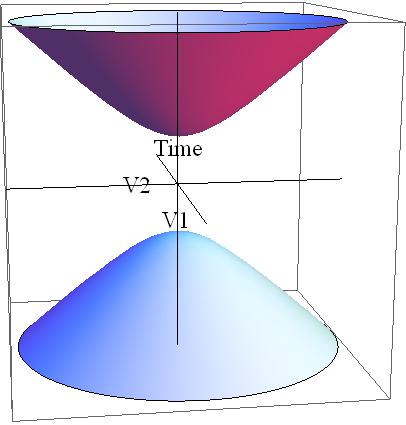} \end{tabular}
\caption{\label{fig:-1} Disconnected sheets of the time-like hyperboloid $V_{\mu}V^{\mu}=$ constant in $\v M_{2+1}$.}
\end{figure}

It was Einstein's discovery \cite{einstein1905} that the invariance of Maxwell's equations found by Lorentz should be imposed also to field sources and particles, giving rise to a structure of space-time sustained by a metric $g=\mbox{diag}\left\{ +1,-1,-1,-1 \right\}$. This is the Minkowski space denoted by $\v M_{3+1}$. Elementary textbooks on particle physics postulate the invariance of four-vector norms under Lorentz transformations in any physical theory, and we proceed in the same manner. We denote a vector that transforms linearly under the Lorentz group as $V_{\mu}$, $\mu=0,1,2,3$ and its contravariant vector as $V^{\mu}=g^{\mu \nu}V_{\nu}$ (summation over repeated indices) such that

\bea
V_{\mu}V^{\mu} = V_0^2 -V_1^2-V_2^2-V_3^2
\label{0.1}
\eea
is an invariant. $V_0$ is the component along the axis of time, and the sign of (\ref{0.1}) determines whether the invariant is time-like ($>0$), space-like ($<0$) or light-like ($=0$). The Lorentz transformations are $4\times 4$ matrices $\Lambda$ with the property

\bea
\Lambda_{\mu \sigma}\Lambda_{\nu \tau} g^{\sigma \tau} = g_{\mu \nu}, \quad V_{\mu}V^{\mu} = V_{\sigma} V^{\tau} \Lambda^{\sigma \mu} \Lambda_{\tau \mu}.
\label{0.2}
\eea
The set of all such matrices forms a six-dimensional abstract surface that has four disconnected components. It is traditionally denoted by O$(1,3)$ (orthogonal group with signature $\left\{+,-,-,- \right\}$). The most common set of transformations in this group is the one connected continuously to the identity; it contains matrices with positive determinant and is denoted by SO$(1,3)$ (special). Using the continuity of the determinant as a function of matrices, we conclude that the components of the group SO$(1,3)$ and O$(1,3)\backslash$SO$(1,3)$ must be disconnected. Each of these two classes also contain two disconnected components, if we recognize that the invariant relation (\ref{0.1}) represents separate sheets of a hyperboloid in space-time, see fig. \ref{fig:-1}. From here it follows that Lorentz transformations cannot map events continuously from one sheet of the hyperboloid (positive time) to the other (negative time). The transformations that preserve the arrow of time are called {\it orthochronous,\ }denoted by SO$^{+}(1,3)$, which is a continuous group by itself. SO$^{+}(1,3)$ contains the identity matrix, together with all the transformations of the form

\bea
\Lambda = \exp \left( i J_{\mu\nu} \theta^{\mu\nu} \right),
\label{0.3}
\eea
where $J_{\mu\nu}$ are the infinitesimal generators of rotations and boosts. The generators $J_{ij}=-J_{ji}$ are true rotations in the plane $x_i$-$x_j$ if $i,j=1,2,3$ while $J_{0i}=-J_{i0}\neq J_{0i}^{\dagger}$ generate the boosts. The six parameters of a transformtation are given by the antisymmetric tensor of 'angles' $\theta^{\mu\nu}$. The reader may consult \cite{greiner1994, barut1986, georgi1999} for a discussion of the Lie bracket related to this group and others.

\begingroup
\begin{table}[b]
\caption{\label{tab:table0}%
The disconnected components of SO$(1,3)$ and SO$(1,2)$
}
\begin{ruledtabular}
\begin{tabular}{l|c|d}
\textrm{ -- }  &
\textrm{Det $=+1$}&
\textrm{Det $=-1$}
\\
\colrule
\textrm{Orthochornous} & SO$^+$ & \textrm{P $\cdot$ SO$^+$} \\
\textrm{Non-orthochronous} &  PT $\cdot$ SO$^+$ &  \textrm{T $\cdot$ SO$^+$}  \\
\end{tabular}
\end{ruledtabular}
\end{table}
\endgroup

In this paper we shall be interested in those transformations that take us (by composition of transformations) from one sheet of the Lorentz group to the others. They are disconnected from the identity and have either negative determinant or time inversion. We shall refer to them as the discrete symmetries of (\ref{0.1}). We have the nomenclature 

\bea
P&=& \left( \begin{array}{cccc}  +1& & & \\ &-1 & & \\ &&-1& \\ &&& -1 \end{array} \right), \quad T= \left( \begin{array}{cccc}  -1& & & \\ &+1 & & \\ &&+1& \\ &&& +1 \end{array} \right), \nonumber \\ PT &=& \left( \begin{array}{cccc}  -1& & & \\ &-1 & & \\ &&-1& \\ &&& -1 \end{array} \right).
\label{0.4}
\eea
See table \ref{tab:table0}. It is important to note that all the elements in one sheet of O$(1,3)$ can be identified with one of the operators in the set $\left\{ \v I_4 , P,T ,PT \right\}$. This set is in fact an abelian group isomorphic to the quotient O$(1,3)/$SO$^{+}(1,3) \cong \v Z_2 \otimes \v Z_2$, which is also known as the Klein group.

In $2+1$ dimensions, we also have four disconnected regions of the group O$(1,2)$ containing the disjoint transformations

\bea
P&=& \left( \begin{array}{ccc}  +1& &  \\ &-1 &  \\ &&+1 \end{array} \right), \quad T= \left( \begin{array}{ccc}  -1& &  \\ &+1 & \\ &&+1 \end{array} \right), \nonumber \\ PT &=& \left( \begin{array}{ccc}  -1& &  \\ &-1 &  \\ &&+1 \end{array} \right).
\label{0.5}
\eea
Note that the parity operator $P$ must have negative determinant and in the $2+1$ dimensional case it reverses the sign of one and only one space component.

\subsection{On the dimensionality of Dirac equations}

Relativistic electrons are described by the Dirac equation \cite{dirac1930}, which contains spin as well as positive and negative energy projections. There are two ways of looking at the origin of this equation. First consider the Lorentz invariant (Klein-Gordon) wave equation

\bea
\left\{ \square + \frac{m^2 c^2}{\hbar^2} \right\} \phi = 0, \quad \square = \frac{ \partial}{\partial x_{\mu}} \frac{\partial }{\partial x^{\mu}} = \frac{1}{c^2}\frac{\partial^2}{\partial t^2} - \nabla^2, \nonumber \\
\label{0.6}
\eea
which merely expresses the energy momentum relation $E^2 = c^2 p^2 + m^2 c^4 $. This equation is of second order in time, and requires the specification of two initial conditions for determining the evolution of waves. Dirac took the 'square root' of (\ref{0.6}) with the purpose of finding a proper relativistic hamiltonian, but such an operation only exists in the space of matrices; they form a Clifford algebra. In simpler units $c=\hbar = 1$ we have the factorization

\bea
\square + m^2 =\left\{ \gamma_{\mu}\frac{\partial}{\partial x_{\mu}} + i m\right\}\left\{\gamma_{\nu}\frac{\partial}{\partial x_{\nu}} - i m \right\}
\label{0.7}
\eea
if and only if the Clifford condition holds

\bea
\{ \gamma_{\mu}, \gamma_{\nu} \} = 2 \v I g_{\mu\nu},
\label{0.8}
\eea
but then a spinorial wave equation should be satisfied:

\bea
\left\{ i \gamma_{\mu}\frac{\partial}{\partial x_{\mu}} - m\right\}\psi = 0.
\label{0.9}
\eea
It is important to recognize here that $\gamma_{\mu}$ is a four-vector of matrices, and that each matrix must be of dimension $4 \times 4$. In fact, a popular representation in terms of Pauli matrices is

\bea
\gamma_0 = \left( \begin{array}{cc} \v 1 & 0 \\ 0 & -\v 1 \end{array} \right), \bfgamma = \left( \begin{array}{cc}  0 & \bfsigma \\ -\bfsigma & 0 \end{array} \right).
\label{1}
\eea
In $2+1$ dimensions the situation is different, since we need only three anticommuting matrices. This time we need only $2\times2$ matrices and they can be represented again in terms of Pauli's $\bfsigma$

\bea
\gamma_0 = \sigma_3, \quad\gamma_1 = i \sigma_2, \quad \gamma_2 = -i \sigma_1.
\label{0.11}
\eea

The implications of dimensionality here are profound, since the spin of the particle in $\v M_{3+1}$ emerges naturally as $\v S = \ahalf \bfsigma$. However, in $\v M_{2+1}$ the spin has only one possible direction, i.e. $S_3 = \ahalf \sigma_3$. In a similar guise, the $4 \times 4$ structure of the Dirac equation in $\v M_{3+1}$ contains information about positive and negative energies or big and small components in the sense of Pauli \cite{bjorken1964}, whereas in $\v M_{2+1}$, $\sigma_1$ and $\sigma_2$ may play such a role without being related to the usual spin. We must warn the reader that effective theories of electrons in two dimensions work with an {\it effective\ }spin generated by lattices, while the true spin of the electron remains as the three-dimensional $\v S$. See section \ref{sec:graphene}.

Yet another way to understand the differences due to dimensionality comes from the representation theory of the groups SO$(1,3)$ and SO$(1,2)$. The Dirac equation is a relation that expresses the invariance of rest mass in the irreducible representation of spin $s=\ahalf$ -- to be precise, the multiplet $(\ahalf,0) \otimes (0,\ahalf)$. We recall here that there is a local isomorphism of our six-dimensional, semi-simple group \cite{greiner1994}

\bea
\mbox{SO}(1,3) \begin{array}{c} _{\cong} \\  \mbox{\scriptsize local}\end{array} \mbox{SU}(2) \otimes \mbox{SU}^*(2).
\label{0.12}
\eea
The lowest irreducible representation of the r.h.s. is a direct product of two sets of Pauli matrices, corresponding to SU$(2)$ and SU$^*(2)$ (the star indicates complex conjugation of the group parameters). Hence the use of $4 \times 4$ $\gamma$ matrices. In contrast, SO$(1,2)$ is a {\it simple\ }and three-dimensional group, requiring only one set of Pauli matrices for the $s=\ahalf$ representation.

\begin{figure}[b]
\begin{tabular}{c}  \includegraphics[width=8cm]{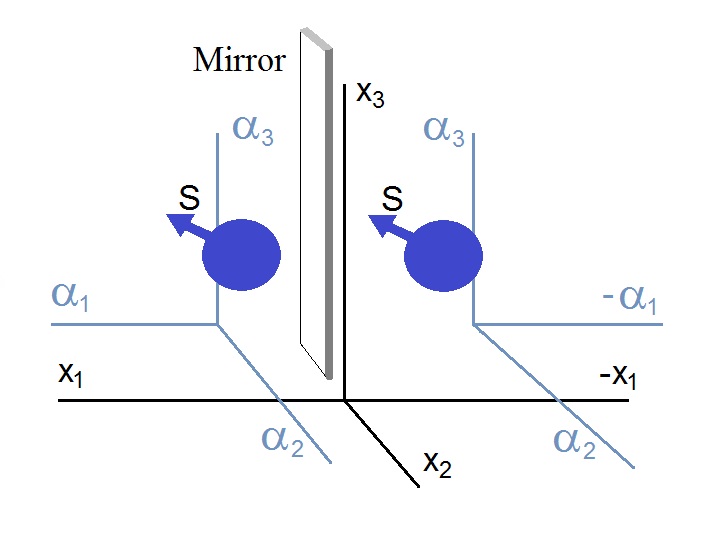} \end{tabular}
\caption{\label{fig:0} Schematic view of parity in $3+1$ dimensions. The wavefunctions corresponding to electrons in opposite sides can be related by a spinorial transformation and an inversion of momenta. The spin is invariant.}
\end{figure}

\begin{figure}[t]
\begin{tabular}{c}  \includegraphics[width=8cm]{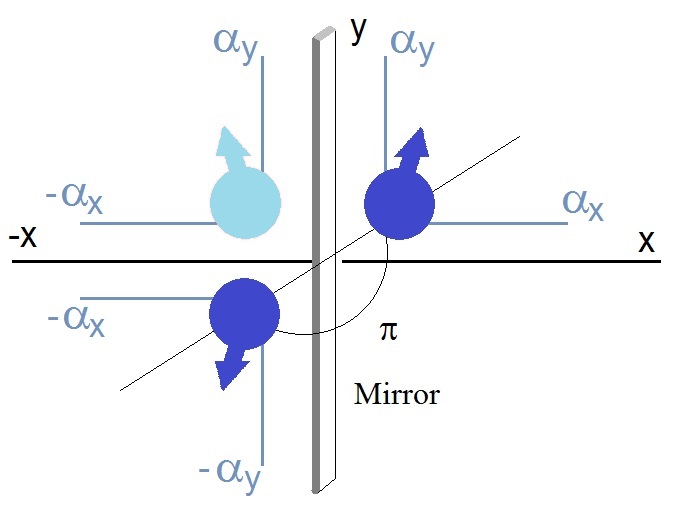} \end{tabular}
\caption{\label{fig:1} (Color online) Parity in $2+1$ dimensions. The dark blue (dark gray) objects represent electrons that can be transformed into each other, whereas the light blue (light gray) object has the same energy spectrum, but obeys a transformed Dirac equation. }
\end{figure}

%
%
%

\section{Parity in low dimensional Dirac equations}

We investigate the difference between $3+1$ and $2+1$ dimensional Dirac equations in regard to discrete transformations. We shall see that the spinorial representations of such objects have important differences due to dimensionality. Among discrete transformations, it is of particular interest to understand parity, as it has been the subject of many discussions in connection with the chiral properties of electrons in two-dimensional materials such as graphene and boron nitride. In our study, the energy-momentum relations must be invariant, although the corresponding equations may vary under discrete transformations. Two diagrams are shown in figures \ref{fig:0} and \ref{fig:1}.

\subsection{A review of parity in 3+1 dimensions}

In order to establish a point of comparison, let us review the transformation properties of the $3+1$ dimensional Dirac equation under parity. This is most easily discussed at the level of first quantization; let $\mu=0,1,2,3$, $i=1,2,3$, and let $\gamma_{\mu}$ be the covariant Dirac matrices in the representation (\ref{1}). In natural units, we write the Dirac equation with momentum $p^{\mu} = i \partial / \partial x^{\mu}=(i\partial/\partial t, -i\nabla)^{ \mbox{\scriptsize T}}$ as

\bea
\left\{ \gamma_{\mu} p^{\mu} - m \right\} \psi(x_{\lambda}) = 0
\label{2}
\eea
or
\bea
\left\{ \gamma_0 p_0-\bfgamma \cdot \v p  - m \right\} \psi(t, \v x) = 0.
\label{3}
\eea
Now we perform the transformation $\v x \mapsto -\v x$, $x_0 \mapsto x_0$ and consequently $\v p \mapsto -\v p$, $p_0 \mapsto p_0$. This results in

\bea
\left\{  \gamma_0 p_0 +\bfgamma \cdot \v p- m \right\} \psi(t, -\v x) = 0.
\label{4}
\eea
We would like to know if there exists a spinorial transformation of $\psi$ such that (\ref{4}) can be transformed back to its original form (\ref{3}), i.e. whether the original wave function and its transformation are described by the same physics. Noting that $\gamma_0 \gamma_i \gamma_0 = - \gamma_i$ and $\gamma_0^2=1$, one has

\bea
\gamma_0\left\{  \gamma_0 p_0 +\bfgamma \cdot \v p - m \right\}\gamma_0 \gamma_0 \psi(t, -\v x) = 0,
\label{5}
\eea
or
\bea
\left\{ \gamma_0 p_0-\bfgamma \cdot \v p  - m \right\}\gamma_0 \psi(t, -\v x) = 0.
\label{6}
\eea
This equation is identical to (\ref{3}), and its solutions $\tilde \psi(t,\v x)$ are such that

\bea
\tilde \psi(t, \v x) = \eta \gamma_0 \psi(t, - \v x),
\label{7}
\eea
where $\eta$ is a global phase factor. This is in fact a transformation law for wavefunctions, and it can be further explored to the level of space-time independent bi-spinors. To this end, let us consider plane waves and spinors in the solution of (\ref{3}) and (\ref{6}). We introduce wave vectors such that $k_{\mu}k^{\mu}= \kappa_{\mu}\kappa^{\mu} = m$ and the normalized bi-spinors $u(k_{\mu}), \tilde u(\kappa_{\mu})$. The wavefunctions read

\bea
\psi(t, \v x) = u(k_{\mu}) \mbox{e}^{-i k^{\nu} x_{\nu}}, \nonumber \\
\tilde \psi(t, \v x) = \tilde u(\kappa_{\mu}) \mbox{e}^{-i \kappa^{\nu} x_{\nu}},
\label{8}
\eea
but in the light of (\ref{7}), we must have the relations

\bea
\bfkappa = - \v k, \quad \kappa_0 = k_0
\label{9}
\eea
and 
\bea
\tilde u(k_0, \v k) = \eta \gamma_0 u(k_0, -\v k).
\label{10}
\eea
This result is in fact quite general, as it can be applied to any superposition of plane waves fulfilling $k_{\mu}k^{\mu}=m$, for which the transformation properties of $u(k_{\mu})$ still hold. In fact, it is customary to use plane wave superpositions with positive ($k_0>0$) and negative ($k_0<0$) energy components of $\psi(t,\v x)$ or their second quantized version \cite{ryder1996}; for the moment we do not need such an expansion.

It is fairly easy to show that other parity transformations (negative determinant) produce similar transformations in spinors. For example, if $x_1 \mapsto - x_1$ with the rest of the components invariant, we obtain

\bea
\kappa_{\mu} = k_{\mu}, \quad \mu \neq 1 \nonumber \\
\kappa_1 = - k_1,
\label{11}
\eea
and
\bea
\tilde u(k_{0},k_{1},k_{2}, k_3) = \eta \gamma_2 \gamma_3 \gamma_0 u(k_{0},-k_{1},k_{2},k_3).
\label{12}
\eea
The spinor transformations (\ref{10}) and (\ref{12}) are mediated by unitary matrices which anticommute with all $\gamma$'s expect for one, and such matrices are built by $\gamma$'s themselves or their products. Is it possible to find similar matrices for problems of different dimensionality? In $2+1$ dimensions, the answer is negative. We shall see this in section \ref{p21}.

\subsubsection{Remarks on PT in $3+1$ dimensions}

Full space-time inversions in $\v M_{3+1}$ are represented by the negative identity matrix. Using the procedures described above, it is easy to show that the PT transformed Dirac equation can be brought back to its original form, and that the wave functions must be related by

\bea
\tilde \psi(x_{\lambda}) = \eta \gamma_5 \psi(-x_{\lambda}),
\label{12.1}
\eea
where $\gamma_5 \equiv i \gamma_0 \gamma_1 \gamma_2 \gamma_3$. It is also worthwhile to recall that the presence of interactions, to the best of our knowledge, respects the CPT symmetry, which includes inversion of charge. In a simplified manner, we may establish this in a Dirac equation with minimal coupling to a gauge field $A_{\mu}$:

\bea
\left\{ \gamma_{\mu} p^{\mu} + e \gamma_{\mu}A^{\mu} - m \right\} \psi(x_{\lambda}) = 0.
\label{12.2}
\eea
If $A^{\mu}$ is a vector, the PT transformation maps $ A^{\mu} \mapsto - A^{\mu}$ and the full equation (\ref{12.2}) is invariant upon the application of $\gamma_5$. On the other hand, if $A^{\mu}$ is a pseudovector, then $ A^{\mu} \mapsto + A^{\mu}$ and the theory is invariant after the application of $\gamma_5$ and the reversal of $e \mapsto -e$. It is also important to remember that charge inversion can be achieved by the successive application of complex conjugation and multiplication by $\gamma_0 \gamma_1 \gamma_3$ (the matrix $\gamma_2$ is complex in the representation we have chosen).

\subsection{Parity in 2+1 dimensions \label{p21}}

Let $\mu=0,1,2$ and $\v x = (x_1,x_2)$. The Dirac equation in $2+1$ dimensions is given now by a $2\times2$ linear differential operator acting on a two-dimensional spinor: 

\bea
\left\{ \sigma_3 p_0 -i\sigma_2 p_1 + i\sigma_1 p_2 - m \right\} \psi(t, \v x) = 0.
\label{13}
\eea
Here, the Dirac matrices are represented by

\bea
\gamma_1 = i \sigma_2, \quad \gamma_2 = -i \sigma_1, \quad \gamma_0 = \sigma_3.
\label{14}
\eea
Now we apply a discrete transformation to (\ref{14}); the space inversion $\v x \mapsto - \v x$ has unit determinant and is irrelevant to our discussion. Let us consider instead $x_1 \mapsto -x_1$ and $x_2 \mapsto x_2$. Our equation (\ref{13}) transforms into

\bea
\left\{ \sigma_3 p_0 + i\sigma_2 p_1 + i\sigma_1 p_2 - m \right\} \psi(t, -x_1, x_2) = 0,
\label{15}
\eea
but this equation cannot be brought to its original form (\ref{13}) by the mere application of unitary operators! Hypothetically, a unitary operator $\Pi$ made of $\gamma$'s that restores the signs in (\ref{15}) must have the properties $\left[ \Pi, \gamma_2 \right]=\left[ \Pi, \gamma_0 \right]=0$ and $\left\{ \Pi, \gamma_1 \right\}=0$. These requirements are impossible to meet in the algebra spanned by all $\gamma$'s and their products, since we have

\bea
\gamma_0 \gamma_1 \gamma_2 = -i \v I, \,
\gamma_0 \gamma_1 = i \gamma_2, \, \gamma_2 \gamma_1 = i \gamma_0, \, \gamma_2 \gamma_0 = i \gamma_1. \nonumber \\
\label{16}
\eea
The first operator commutes with everything, while the other operators in (\ref{16}) applied to (\ref{15}) would produce two sign flips (positive determinant). A similar situation occurs when we try to introduce complex conjugation as a possible transformation; we have

\bea
(\gamma_0 p_0)^{*} = - \gamma_0 p_0, \quad (\gamma_1 p_1)^{*} = - \gamma_1 p_1, \quad (\gamma_2 p_2)^{*} = + \gamma_2 p_2,\nonumber \\
\label{17}
\eea
and two sign flips would occur again in (\ref{15}). With this, we conclude that the wavefunctions $\psi(t,-x_1,x_2)$ and $\psi(t,x_1,x_2)$ cannot be transformed into each other, although they may satisfy the same energy-momentum relation $k_{\mu}k^{\mu}=m$ when expanded in plane waves.

In a theory of many fermions (for example, the second quantization of the theory above) it seems necessary to introduce at least two flavors that account for all possible solutions of the energy-momentum relation but whose equations are inequivalent. We shall see in section \ref{sec:graphene} that this is exactly the case for some two-dimensional systems in condensed matter. 

Returning to first quantization and the Dirac equation, we point out that a happy accident occurs in the absence of mass. The Dirac operator becomes $\gamma_{\mu}p^{\mu}$; although this operator is not invariant under $x_1 \mapsto - x_1$, it turns out that this transformation can be continuously related with a full space-time inversion: the relation

\bea
\left\{ \gamma_0 p_0 + \gamma_1 p_1 - \gamma_2 p_2  \right\} \psi(t,-x_1,x_2)=0
\label{18}
\eea  
can be transformed by applying $-\gamma_1$ from the left

\bea
\left\{   \gamma_0 p_0  - \gamma_1 p_1 - \gamma_2 p_2\right\} \gamma_1 \psi(t,-x_1,x_2)=0
\label{19}
\eea
which is the sought result. This shows that the massless Dirac equation {\it is invariant\ }under $x_1 \mapsto - x_1$ and the solutions are related by

\bea
\tilde \psi(t,x_1,x_2) = \eta \gamma_1 \psi(t,-x_1, x_2)
\label{20}
\eea
where $\eta$ is again a phase factor, including signs. We note that the transformation is now mediated by $\gamma_1$, whereas in the $3+1$ dimensional case the matrix was $\gamma_0$.

\subsubsection{Hamiltonian formulation in $2+1$ dimensions}

The previous results are not too different when we bring the Dirac equation to a hamiltonian form. Here of course, time reversal transformations without energy sign reversal require antilinear operators. It also happens that parity-transformed hamiltonians may have the same spectrum, and indeed $E=\pm \sqrt{p^2 + m^2}$ is invariant under parity. With the traditional notation $\alpha_1=\sigma_1, \alpha_2=\sigma_2, \beta= \sigma_3$ we have the Schr\"odinger equation

\bea
\left\{ \bfalpha \cdot \v p + m \beta \right\} \psi(t, \v x) = i \frac{\partial \psi(t, \v x)}{\partial t}.
\label{21}
\eea
Although the operator

\bea
H^{2+1} = \bfalpha \cdot \v p + m \beta
\label{22}
\eea
is not a parity invariant, the spectrum is invariant. This implies that the eigenfunctions are divided at least in two classes (as we saw previously), producing degeneracy when both the original and the transformed hamiltonian belong to the same theory. 

We examine again the parity transformations at the level of (\ref{21}) and its stationary version. Take $\psi(t,\v x) = \mbox{e}^{-i E t} \phi(\v x)$ and perform the transformation $x_1 \mapsto -x_1$, $x_2 \mapsto x_2$ to find

\bea
\left\{-\sigma_1 p_1+ \sigma_2 p_2 + m\sigma_3 \right\} \phi(-x_1,x_2) = E \phi(-x_1,x_2).\nonumber \\
\label{23}
\eea 
Here complex conjugation pays off (but not in the full time-dependent solution!), as it leads to

\bea
\left\{ \sigma_1 p_1+ \sigma_2 p_2 + m\sigma_3 \right\} \phi^{*}(-x_1,x_2) = E \phi^{*}(-x_1,x_2). \nonumber \\
\label{24}
\eea
For this reason $\phi^{*}(-x_1,x_2)$ and $\phi(x_1,x_2)$ have the same energy, but it is left to see whether these solutions are independent or not with respect to their spinorial part. Once again we use a single plane wave to see that if

\bea
\phi(\v x) = u(\v k) \mbox{e}^{i \v k \cdot x}
\label{25}
\eea
then
\bea
\left\{ \bfsigma \cdot \v k + m \sigma_3 \right\} u(\v k) = E u(\v k),
\label{26}
\eea
with its complex conjugate given by
\bea
\left\{ \sigma_1 k_1 -\sigma_2 k_2  + m \sigma_3 \right\} u^{*}(\v k) = E u^{*}(\v k).
\label{27}
\eea
Now we must have that $u$ and $u^{*}$ are independent, for the proportionality $u \propto u^{*}$ leads to the contradictory relation $k_2 \sigma_2 u = 0$ by the combination of (\ref{26}) and (\ref{27}). So $u$ is necessarily complex, and the spinors corresponding to opposite parities and equal energies are independent ($k_2=0$ is possible, but reduces effectively the problem to one dimension, and is not of interest).

In conclusion, in $2+1$ dimensions only the {\it stationary\ }solutions of opposite parity can be related by a transformation, which turns out to be a complex conjugation, involving thus antiunitary operators. The complex character of the wavefunction and its spinorial part makes $\phi(x_1, x_2)$ and $\phi^*(-x_1,x_2)$ independent.

\subsubsection{Remarks on PT in $2+1$ dimensions}

Full space-time inversion produces three sign flips in (\ref{13}) and is therefore continuously connected to the $x_1 \mapsto - x_1$ transformation. For this reason, the functions $\psi(-x_{\mu})$ and $\psi(x_{\mu})$ cannot be transformed into each other. How about the functions $\psi(-t,-x_1,x_2)$ and $\psi(t,x_1,x_2)$? This PT transformation can be reproduced by the application of the matrix $\gamma_2$ or by complex conjugation. With this we can show that the functions $\psi(t,x_1,x_2)$, $\psi^{*}(-t,-x_1,x_2)$ and $\gamma_2\psi(-t,-x_1,x_2)$ can be transformed into each other, fulfilling the glorified CPT invariance. At the hamiltonian level we can easily show that the transformation involves energy inversion; the reversed parity equation

\bea
\left\{ -\sigma_1 p_1 + \sigma_2 p_2 + m \sigma_3 \right\} \phi(-x_1,x_2) = E \phi(-x_1,x_2)  \nonumber \\
\label{28}
\eea
is transformed now to

\bea
\left\{ \sigma_1 p_1 + \sigma_2 p_2 + m \sigma_3 \right\} \sigma_1 \phi(-x_1,x_2) = -E\left[ \sigma_1 \phi(-x_1,x_2) \right] \nonumber \\
\label{29}
\eea
after multiplying by $-\sigma_1$ from the left. This can be resumed as follows: a function $\tilde \phi$ of positive energy $E$ can be expressed in terms of negative energy solutions in the form

\bea
\tilde \phi_E (x_1,x_2) = \eta \sigma_1 \phi_{-E}(-x_1,x_2)
\label{30}
\eea
where $\eta$ is again a global phase factor. With this we show that the symmetric spectrum of this theory (about the point $E=0$) is related to transformations under P alone.


\section{Graphene and Boron Nitride: effective theories in flat sheets \label{sec:graphene}}

\begin{figure}[t]
\begin{tabular}{c}  \includegraphics[width=5cm]{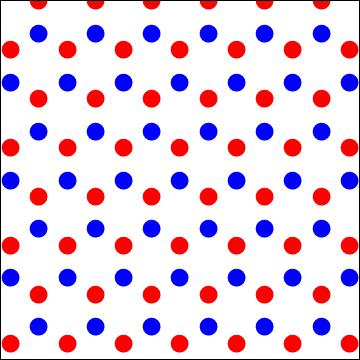} \end{tabular}
\caption{\label{fig:2} (Color online) An hexagonal lattice formed by two interpenetrating triangular sublattices in blue and red. }
\end{figure}

\begin{figure}[h!]
\begin{tabular}{c}  \includegraphics[width=6cm]{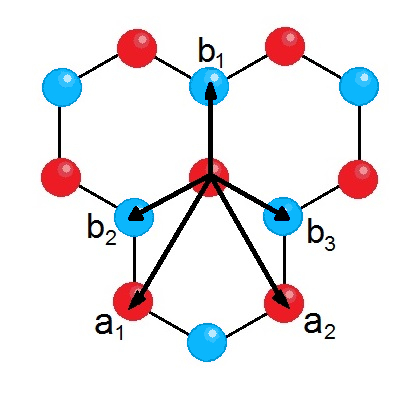} \end{tabular}
\caption{\label{fig:3} (Color online) Fundamental cell of the hexagonal lattices and primitive vectors. Blue and red sites (dark and light gray) may represent different types of atoms. }
\end{figure}

It has been noted in the literature of condensed matter physics \cite{katsnelson2007, geim2007}, that electrons in hexagonal lattices (see figure \ref{fig:2}) can be described by effective $2+1$ dimensional Dirac equations. It turns out that there are inequivalent conical points at the edges of the first Brillouin zone (in this case an hexagon) of the honeycomb lattice, where the dispersion relations of propagating waves resemble a relativistic energy-momentum relation \cite{semenoff1984}:

\bea
E=\epsilon - \epsilon_0 \approx \pm \sqrt{\Delta^2 (\v k \pm \v k_D)^2 + m^2}
\label{31}
\eea 
where $\v k$ is the Bloch momentum of a wave in the crystal, $\v k_D$ is the point of maximal approach of positive and negative surfaces (the famous Dirac points \cite{haldane2008, bittner2010, gomes2012}), $\Delta$ is the nearest-neighbor coupling in the corresponding tight-binding model (in condensed matter physics $\Delta$ is related to the Fermi velocity), $\epsilon_0$ is the center of the lowest energy band and $m$ is the difference between binding energies of atoms at each triangular sublattice (examples with two species include boron nitride, while $m=0$ describes graphene.) In addition to this appealing dispersion relation, one also has an effective spin given by the probability of being in sites of type A or B (see figure \ref{fig:3}). Incidentally, this spin is represented by $\bfsigma$ matrices, in full correspondence with our previous considerations of Dirac equations in $2+1$ dimensions.

\begin{figure}[t]
\begin{tabular}{c}  \includegraphics[width=6cm]{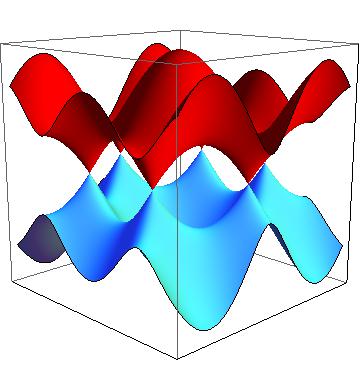} \end{tabular}
\caption{\label{fig:4} Dispersion relation in the reciprocal honeycomb lattice. Six conical points can be distiguished. Opposite points are inequivalent.}
\end{figure}

\begin{figure}[h!]
\begin{tabular}{c}  \includegraphics[width=6cm]{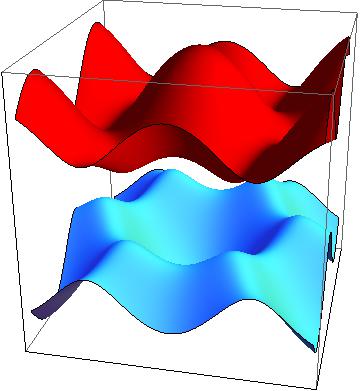} \end{tabular}
\caption{\label{fig:5} Dispersion relation for the massive case: the gap between the blue (upper) and red (lower) bands is originated by a difference of on-site energies between A and B. }
\end{figure}

\subsection{Parity in effective theories with two fermions \label{p2fermions}}

In such effective theories we have two types of Dirac equations fulfilling the dispersion relation (\ref{31}):

\bea
\left\{ \gamma_0 p_0 - \gamma_1 p_1 - \gamma_2 p_2 - m \right\} \psi^{+}=0,\nonumber \\
\left\{ \gamma_0 p_0 + \gamma_1 p_1 - \gamma_2 p_2 - m \right\} \psi^{-}=0, 
\label{32}
\eea
where $\v p$ is now the momentum around the point $\pm \v k_D$, with eigenvalues $\Delta (\v k \mp \v k_D)$. There are no translations in the reciprocal triangular sublattice that could take us from $\v k_D$ to $-\v k_D$, and we have seen previously that the wavefunctions cannot be transformed into each other. The full theory, however, is invariant under the interchange $+ \leftrightarrow -$. Schematically, we may describe both relations in (\ref{32}) by a single bi-spinorial equation:

\bea
\left(\begin{array}{cc} \gamma_{\mu} p^{\mu} - m & 0 \\ 0 &  -\gamma_1 (\gamma_{\mu} p^{\mu})\gamma_1 - m \end{array} \right)\left(\begin{array}{c} \psi^{+}(x_{\lambda})  \\  \psi^{-}(x_{\lambda}) \end{array} \right)=0. \nonumber \\
\label{33}
\eea
We can perform now a parity operation to finally understand why these electrons obey a chiral theory: if $x_1 \mapsto -x_1$ and $p_1 \mapsto -p_1$, the roles of $\pm$ will be interchanged, i.e.

\bea
\left(\begin{array}{cc}  -\gamma_1 (\gamma_{\mu} p^{\mu})\gamma_1 - m & 0 \\ 0 & \gamma_{\mu} p^{\mu} - m \end{array} \right)\left(\begin{array}{c} \psi^{+}(t,-x_1,x_2)  \\  \psi^{-}(t,-x_1,x_2) \end{array} \right)=0. \nonumber \\
\label{34}
\eea
The complete theory is invariant if we apply the $4 \times 4$ swapping operator

\bea
\Gamma \equiv \left(\begin{array}{cc}  0 & \v I_2 \\  \v I_2 & 0 \end{array} \right)
\label{35}
\eea
to the bi-spinor 

\bea
\Psi(t,-x_1,x_2) \equiv \left(\begin{array}{c} \psi^{+}(t,-x_1,x_2)  \\  \psi^{-}(t,-x_1,x_2) \end{array} \right)
\label{36}
\eea
and to the augmented Dirac operator (as a similarity transformation)

\bea
D(p_0,p_1,p_2) \equiv \left(\begin{array}{cc} \gamma_{\mu} p^{\mu} - m & 0 \\ 0 &  -\gamma_1 (\gamma_{\mu} p^{\mu})\gamma_1 - m \end{array} \right).\nonumber \\
\label{37}
\eea
We explain the invariance as follows. By virtue of the relations $\Gamma^2 = \v I_4 $, $\Gamma D(p_0,-p_1,p_2) \Gamma = D(p_0,p_1,p_2)$, we have that if

\bea
D(p_0,p_1,p_2) \Psi(t,x_1,x_2) = 0,
\label{38}
\eea
then
\bea
D(p_0,-p_1,p_2) \Psi(t,-x_1,x_2) = 0
\label{39}
\eea
and
\bea
D(p_0,p_1,p_2) \Gamma \Psi(t,-x_1,x_2)=0.
\label{40}
\eea
The exchange of $\pm$ does the trick. At the level of Hamiltonians the theory is also invariant: defining

\bea
 \hcal(\v p) \equiv \left(\begin{array}{cc} \bfalpha \cdot \v p + m \beta  & 0 \\ 0 & \sigma_2 (\bfalpha \cdot \v p) \sigma_2  + m \beta \end{array} \right)
\label{41}
\eea
with stationary functions

\bea
\Psi(t, \v x) = \mbox{e}^{-iEt} \Phi(\v x),
\label{42}
\eea
we obtain $\Gamma \hcal (-p_1,p_2)\Gamma = \hcal(p_1,p_2)$ and

\bea
\hcal(p_1,p_2) \Gamma \Phi(-x_1,x_2) = E \left[\Gamma \Phi(-x_1,x_2)\right]
\label{43}
\eea
as expected. There is nothing artificial about this procedure, if we regard the theory as made of two types of fermions with equal probability of existence. However, this interpretation leads invariably to more than one particle in the hexagonal sheet (in fact, many of them). This makes sense only in a second-quantized scheme of solid state physics.

It is thus natural to ask whether a single-particle formulation may have a similar chiral symmetry. The answer is positive, if we take into account the {\it complete\ }spectrum of the theory, without the conical approximations (\ref{31}) related to effective Dirac equations. Furthermore, it also holds that even without the conical approximation of the dipersion relations, the theory still has a spinorial formulation (spin up and down are A and B) where the effective matrices can be defined solely by the geometry of the lattice \cite{sadurni2010}. We shall play with this formulation in what follows, with the aim of extracting once more the spinorial representations of discrete transformations, but this time in the context of crystals.

\subsection{Parity in a complete tight-binding model with one fermion}

\subsubsection{The general model with Dirac matrices}

The full tight-binding model can be constructed starting from very simple considerations \cite{wallace1947}. For the sake of clarity we discuss it here along the lines indicated in \cite{sadurni2010}. The honeycomb lattice is defined by two interpenetrating triangular sublattices with primitive vectors $\v a_1 = (a /2)(-\v i - \sqrt{3}\v j), \v a_2 = (a /2)(\v i - \sqrt{3}\v j)$. Each point has three nearest neighbors; the origin is connected to such sites by the vectors $\v b_1 = (a/\sqrt{3}) \v j, \v b_2=(a/2)(-\v i - (1/\sqrt{3})\v j), \v b_3 = -\v b_1 - \v b_2$. See figures \ref{fig:2} and \ref{fig:3}. We can label the atomic states\footnote{These are isolated atomic states, but it is more appropriate to use Wannier functions. For maximally localized states, see \cite{marzari1997}.} by site vectors $\v A$ and $\v B$ corresponding to each sublattice, i.e. $|\v A\>$ and $|\v B\>$. They are linear combinations of $\v a_1, \v a_2$ with integer coefficients and the term $\v b_1$ is added in the case of $\v B$. The most common way to write a nearest-neighbor tight-binding model in first quantization is the following:

\bea
H &=& \Delta \sum_{\v A, i=1,2,3} | \v A \> \< \v A + \v b_i | + h.c. \nonumber \\
&+& E_A \sum_{\v A} | \v A \> \< \v A | +  E_B \sum_{\v B} | \v B \> \< \v B |,  
\label{44}
\eea
where $E_A$ and $E_B$ are the binding energies of atoms in lattice A and B respectively. A more convenient way to write this hamiltonian can be achieved by introducing translation operators and some definitions. The goal is to express (\ref{44}) in a way similar to a Dirac hamiltonian. We need Dirac matrices $\bfalpha = \bfsigma$, and we may define them in terms of localized states

\bea
\sigma_{1} = \sum_{\v A} \left[| \v A \> \< \v A + \v  b_1 | +  | \v A + \v b_1 \> \< \v A | \right], \nonumber  \\
\sigma_{2} = -i \sum_{\v A} \left[ | \v A \> \< \v A + \v  b_1 | -  | \v A + \v b_1 \> \< \v A |\right], \nonumber \\
\sigma_{3} = \sum_{\v A} \left[| \v A \> \< \v A | -  | \v A + \v b_1 \> \< \v A + \v b_1 | \right],
\label{45}
\eea
which satisfy the SU$(2)$ algebra $\left[ \sigma_i, \sigma_j \right]=2i \epsilon_{ijk}\sigma_k$ and the Clifford condition $\left\{ \sigma_i, \sigma_j \right\} = 2 \v I_2 \delta_{ij}$. Similarly, we define operators analogous to momenta in the form

\bea
P_1 = \frac{\Delta}{2} \sum_{\v A, i} |\v A +\v b_i \> \< \v A + \v b_1 | + |\v A +\v b_i  - \v b_1 \> \< \v A | + \mbox{h.c.} , \nonumber \\
P_2 = \frac{\Delta}{2i} \sum_{\v A, i} |\v A +\v b_i \> \< \v A + \v b_1 | + |\v A +\v b_i  - \v b_1 \> \< \v A | + \mbox{h.c.} \nonumber \\
\label{46}
\eea
It is important to note that $P_1$ and $P_2$ are made of translation operators $T_i = \exp \left( i \v a_i \cdot \v p \right)$ connecting sites of the same subtriangular lattice, i.e.
\bea
P_1 &=& \frac{\Delta}{2} \left[ 2 \v I + T_1 + T_1^{\dagger} + T_2 + T_2^{\dagger} \right] ,\nonumber \\
P_2 &=& \frac{\Delta}{2i} \left[ T_1 - T_1^{\dagger} + T_2 - T_2^{\dagger} \right].
\label{47}
\eea 
With these identifications, we finally arrive at the hamiltonian

\bea
H = \bfalpha \cdot \v P + m \beta + \epsilon_0
\label{48}
\eea
where $m=(E_A-E_B)/2$ and $\epsilon_0 = (E_A+E_B)/2$. Here, we are only one step away from an effective Dirac theory, since the expansions of the exponentials $T_i$ in $P_1$ and $P_2$ around $\Delta \v k_D$, yield linear expressions in $p_1$ and $p_2$ respectively (conical points). However, the full theory with hamiltonian (\ref{48}) has eigenvalues

\bea
\epsilon = \epsilon_0 \pm \sqrt{ \Delta^2 | 1 + \mbox{e}^{i \v k \cdot \v a_1} + \mbox{e}^{i \v k \cdot \v a_2} |^2 + m^2}, 
\label{49}
\eea
which can be computed using Bloch waves $\sum_{\v A} \mbox{e}^{i \v k \cdot \v A} \<\v x | \v A\>$ in each spinor component. Such spinors diagonalize the following $2 \times 2$ blocks in the hamiltonian

\bea
\left(\begin{array}{cc} \epsilon_0 + m& \Delta \left[ 1 + \mbox{e}^{i \v k \cdot \v a_1} + \mbox{e}^{i \v k \cdot \v a_2} \right] \\ \Delta \left[1 + \mbox{e}^{-i \v k \cdot \v a_1} + \mbox{e}^{-i \v k \cdot \v a_2} \right] & \epsilon_0 - m \end{array} \right). \nonumber \\
\label{49.1}
\eea

\subsubsection{The new $\v P$ as a pseudovector}

Now we are ready to discuss the parity transformation $x_1 \mapsto -x_1, x_2 \mapsto x_2$. We have $p_1 \mapsto - p_1$, $p_2 \mapsto p_2$, but in view of the property $\v a_1 \cdot \v p \mapsto \v a_2 \cdot \v p$ and vice versa, the translation operators are now mapped into each other

\bea
T_1  \mapsto  T_2, \quad T_2 \mapsto T_1,
\label{50}
\eea
leading to a pseudovectorial $\v P$:

\bea
P_1 \mapsto P_1 \quad P_2 \mapsto P_2.
\label{51}
\eea
With these relations, the invariance of the full hamiltonian (\ref{48}) is ensured. 

Incidentally, the Dirac point at $\v k_D = (4 \pi / 3 a)\v i $ is mapped to $- (4 \pi / 3 a)\v i$, which is the inequivalent Dirac point at the opposite vertex. However, both vertices are contained in our single particle theory and its invariance is again confirmed. As to the wavefunctions, the spatiotemporal part is given by Bloch waves and only a change $k_1 \mapsto -k_1$ is needed. The spinorial part remains invariant.

\begin{figure}[b]
\begin{tabular}{c}  \includegraphics[width=9.0cm]{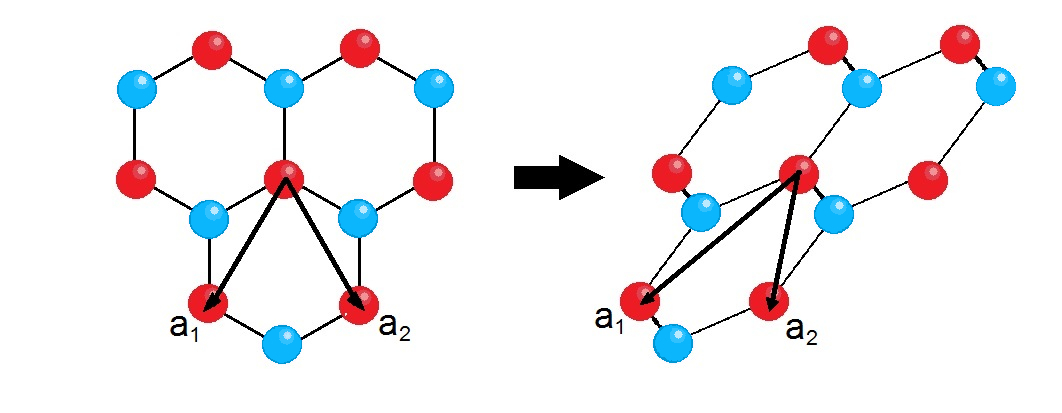} \end{tabular}
\caption{\label{fig:9} Parity symmetry breaking by sheet deformation. The bonds represented by vectors $\v a_1$ and $\v a_2$ have different lengths and different couplings.}
\end{figure}

\begin{figure}[t]
\begin{tabular}{c}  \includegraphics[width=5cm]{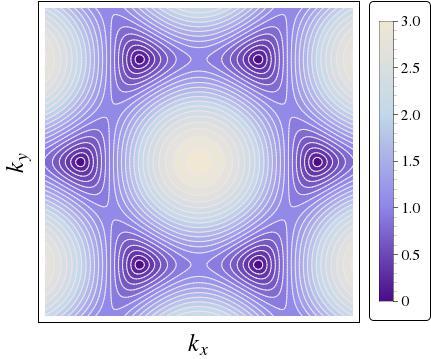} \end{tabular}
\caption{\label{fig:10} Upper view of the dispersion relation surface, showing a symmetric hexagon. $\Delta=1$. }
\end{figure}

\begin{figure}[t]
\begin{tabular}{c}  \includegraphics[width=5cm]{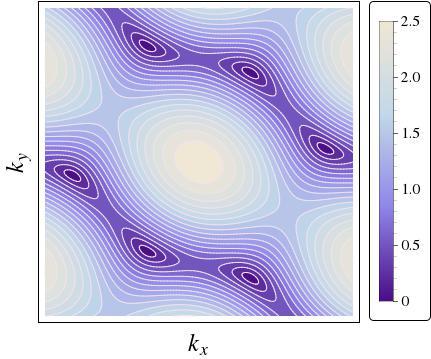} \end{tabular}
\caption{\label{fig:11} Upper view of an asymmetric dispersion relation induced by sheet deformation. $\Delta_1=1, \Delta_2 = 1/2$.}
\end{figure}

\subsection{Discrete symmetry breaking}

There are several ways to introduce interactions which violate discrete symmetries. In particle physics we may quote famous examples \cite{garwin1957, lees2012} in which a partial discrete symmetry is violated, such as parity (weak interactions) or time reversal and charge conjugation (CP violation). There are even more exotic proposals \cite{colladay1997} that suggest CPT violation as an effect that emerges due to novel theories. In this paper we restrict ourselves to the importance of dimensionality and its implications in effective theories on the lattice. A most fascinating consequence of reduced dimensionality is the so-called chiral anomaly\cite{niemi1983, quackenbush1989}, which indeed is represented by two types of electrons in hexagonal lattices that suffer transitions from one type to the other (interpreted as tunneling) due to quantum corrections. In connection with explicit symmetry breaking, i.e. at the loevel of the hamiltonian, it is easy to see that lattice deformations do the job in two different forms: 1) by breaking A$\leftrightarrow $B invariance, leading to the appearance of an effective mass as we saw previously and 2) by introducing bond asymmetries (see fig. \ref{fig:9}), e.g. by applying a shear.

\subsubsection{Two fermions}

In section \ref{p2fermions} we saw that the hamiltonian of the theory could be expressed by an augmented operator $\hcal$. The exchange invariance of the theory can be broken easily by introducing a non-diagonal operator in the space of spinors $\psi^{\pm}$. An example of such an interaction which does not commute with $\Gamma$ can be proposed to be proportional to

\bea
\bar \Gamma = \left( \begin{array}{cc} 0  &  -i \v I_2  \\ i\v I_2 & 0  \end{array}\right).
\label{52}
\eea
Evidently, this leads to transitions between the two species. Diagonal terms which do not commute with $\Gamma$ can be conceived as well, but they do not correspond to a coupling between the two inequivalent Dirac points.

\subsubsection{Full band theory with one fermion}

A very general way to break the invariance of $\hcal$ under parity is by the introduction of vectorial interactions. When such potentials are external, i.e. not dynamical variables of the world, their transformation properties are determined solely by the coordinates. For example, if

\bea
\left\{ \gamma_0 p_0 - \gamma_1 P_1 - \gamma_2 P_2 - m  + V_{ \mbox{\scriptsize int}} \right\} \psi = 0
\label{53}
\eea
then $V_{ \mbox{\scriptsize int}} \equiv \gamma_{\mu} V^{\mu}$ would do the job, as long as $V^{\mu}$ transforms as a vector under parity (remember that $\v P$ is a pseudovector).

Another way to break parity symmetry is by introducing complex couplings $\Delta$, such as those used to simulate gauge fields \cite{uehlinger2013}, in particular external magnetic fields. The asymmetry in the lattice bonds can be introduced generally as

\bea
P_1 &=& \frac{1}{2} \left[ 2 \Delta_0 + \Delta_1 T_1 + \Delta_1 T_2 \right] + \mbox{h.c.} ,\nonumber \\
P_2 &=& \frac{1}{2i} \left[ \Delta_1 T_1 + \Delta_2 T_2 \right] + \mbox{h.c.}.
\label{54}
\eea
where $\Delta_i$ are complex. If $\Delta_1 \neq \Delta_2$, then the exchange $\v a_1 \leftrightarrow \v a_2$ is no longer a symmetry of the hamiltonian. Generically, there is no way in which the application of operators depending on $\gamma$ matrices may restore the symmetry, and the theory is not invariant. There are two cases to be distinguished: When only the phases of $\Delta_1, \Delta_2$ are different, we recognize that they can be redefined by the application of unitary transformations forming a gauge group U$(1)$. This represents indeed a magnetic field. When the moduli are different, i.e. $|\Delta_1| \neq |\Delta_2|$ then the bonds mediated by the vectors $\v b_2$ and $\v b_3$ are different, a type of asymmetry that can be introduced by a constant deformation that modifies the fundamental cell, but not the periodicity of the medium. The overall effect in such theories amounts to a modification of the dispersion relation. This effect has been extensively investigated \cite{montambaux2009} with the pupose of translating and merging inequivalent Dirac points. A comparison of energy surfaces is given in figures \ref{fig:10} and \ref{fig:11}.

\begin{figure}[t]
\begin{tabular}{c}  \includegraphics[width=6cm]{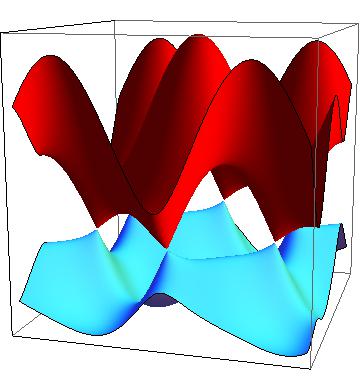} \end{tabular}
\caption{\label{fig:6} Asymmetric bands produced by the introduction of next-to-nearest neighbor interactions. The upper and lower surfaces are different.}
\end{figure}

\begin{figure}[t]
\begin{tabular}{c}  \includegraphics[width=5cm]{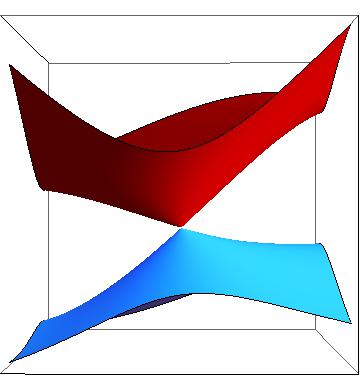} \end{tabular}
\caption{\label{fig:7} Asymmetric bands induced by second neighbors, visualized around conical points. Although the complete system must be CPT symmetric, the effective theory of the electron is not. }
\end{figure}

\begin{figure}[t]
\begin{tabular}{c}  \includegraphics[width=5cm]{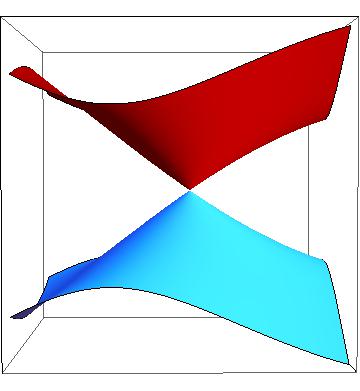} \end{tabular}
\caption{\label{fig:8} Usual cones with up-down symmetry. Compare with fig. \ref{fig:7}.}
\end{figure}

Another interesting possibility comes in the form of mutliple neighbor couplings. It turns out that their presence can break the symmetry between upper and lower bands around conical points, indicating that the effective CPT symmetry of the theory (the one that relates particles with antiparticles or electrons with holes) can be broken. The explicit way to achieve this is by adding terms to $\hcal$ as follows 

\bea
\hcal = \epsilon_0 + m \sigma_3 + \bfalpha \cdot \v P + \bar \Delta (T_1 + T_2 + T_1 T_2^{\dagger} + \mbox{h.c.}).\nonumber \\
\label{55}
\eea
In this expression, the last term does not contain Dirac matrices, and it couples the six second neighbors of each site by connecting them through the vectors $\pm a_1, \pm a_2, \pm (a_1 - a_2)$. The constant $\bar \Delta$ modulates the interaction. The resulting dispersion relation and a comparison between energy cones is given in figures \ref{fig:6}, \ref{fig:7} and \ref{fig:8}. Here we should note that a parity transformation leaves such terms invariant (this is again the exchange $\v a_1 \leftrightarrow \v a_2$), but the application of $PT$ at the level of the Dirac equation

\bea
&& \left\{ \gamma_0p_0 - \bfgamma \cdot \v P - m -\bar \Delta \gamma_0 (T_1 + T_2 + T_1 T_2^{\dagger} + \mbox{h.c.}) \right\} \nonumber \\ 
&\times &  \psi(t,x_1,x_2) = 0
\label{56}
\eea
 reveals that

\bea
\gamma_2 \left\{ -\gamma_0 p_0 + \gamma_1 P_1 - \gamma_2 P_2 - m \right\}\psi(-t,-x_1,x_2) - \nonumber \\ 
  \bar \Delta \gamma_2 \gamma_0 (T_1 + T_2 + T_1 T_2^{\dagger} + \mbox{h.c.}) \psi(-t,-x_1,x_2) = 0
\label{57}
\eea
or put another way

\bea
&& \left\{ \gamma_0p_0 - \bfgamma \cdot \v P - m + \bar \Delta \gamma_0 (T_1 + T_2 + T_1 T_2^{\dagger} + \mbox{h.c.}) \right\} \nonumber \\ 
&\times & \gamma_2 \psi(-t,-x_1,x_2) = 0.
\label{58}
\eea
This equation is not equivalent to (\ref{56}), and the only possible way to restore the sign of the last term is by coupling inversion $\bar \Delta \mapsto - \bar \Delta$. In a world where the actors are transformed but the stage is fixed, such a coupling inversion is not allowed and the dispersion relation must have an up-down asymmetry. Obviously, when the stage is also reversed, we recover CPT invariance of our complete world.

\section{Discussion}

The role of discrete symmetries in both particle physics and condensed matter systems should not be underestimated. In this paper we have reviewed the subject at the level of the Dirac equation in first quantization. It is important to mention that a frequent approach to symmetries in quantum field theory comes from the invariance of the action that generates the Euler-Lagrange equations, including the Dirac equation. Invariance of the action leads indeed to invariance of the theory, but the converse is not necessarily true; the subtleties of this and other properties arising in a second-quantized scheme have been left aside for the sake of a simple treatment. We encourage our readers with particle physics inclinations to consult references \cite{beringer2012} with respect to state-of-the-art CPT invariance tests.

As to the honeycomb lattice, there is a clear message arising from our results: lattice deformations and long range interactions constitute a source of asymmetry that can be used to our favor as a testbed for new effects. However, we must warn the reader that the validity of conical approximations in graphene has been experimentally established for energies in the vicinity of the band center. Thus, the effects arising due to a full-band theory may be visible in other honeycomb realizations. The so-called artificial graphene \cite{kalesaki2014} is worthy of attention.


\begin{acknowledgments}
E. S. and E. R.-M. would like to express their gratitude to CONACyT for financial support under project CB2012-180585. 
\end{acknowledgments}

\nocite{*}



\providecommand{\noopsort}[1]{}\providecommand{\singleletter}[1]{#1}%

\end{document}